\documentclass[twocolumn]{aa} 

\pdfoutput=1
\usepackage{graphicx}
\usepackage{txfonts}
\usepackage{color}
\usepackage{amsmath}
\usepackage{amssymb}
\usepackage{rotating}
\usepackage{verbatim}
\usepackage{subfig}
\usepackage{natbib}
\usepackage{booktabs}
\usepackage{longtable}
\usepackage{mathtools,cuted}
\begin{document}

\title{Planetary Radio Interferometry and Doppler Experiment (PRIDE) Technique: a Test Case of the Mars Express Phobos Fly-by.}
\subtitle{2. Doppler tracking: Formulation of observed and computed values, and noise budget.}

   \author{T. M. Bocanegra-Baham\'{o}n
          \inst{1,2,3}
          \and
          G. Molera\,Calv\'{e}s
          \inst{1,4}
          \and 
          L.I. Gurvits
          \inst{1,2}
	  \and
          D.A. Duev
          \inst{5}
          \and
          S.V. Pogrebenko
          \inst{1}
          \and
          G. Cim\`{o}
          \inst{1,6}
          \and
          D. Dirkx
          \inst{2}
          \and
          P. Rosenblatt
          \inst{7}          
          }

   \institute{Joint Institute for VLBI ERIC, P.O. Box 2, 7990 AA Dwingeloo, The Netherlands.\\
              \email{bocanegra@jive.eu}
         \and
             Department of Astrodynamics and Space Missions, Delft University of Technology, 2629 HS Delft, The Netherlands.\\
         \and
             Shanghai Astronomical Observatory, 80 Nandan Road, Shanghai 200030, China. \\
 	 \and 
	     Finnish Geospatial Research Institute, National Land Survey of Finland, Geodenterinne 2, 02430 Masala, Finland.\\
	 \and
             Department of Astronomy, California Institute of Technology, 1200 E California Blvd, Pasadena, CA 91125, USA. \\
      \and 
     Netherlands Institute for Radio Astronomy, P.O. Box 2, 7990 AA Dwingeloo, The Netherlands. \\             
     \and 
             Royal Observatory of Belgium, Ringlaan 3, 1180 Brussels, Belgium.}

 \date{Received July 6, 2017; accepted September 7, 2017}

 
  \abstract
   {Closed-loop Doppler data obtained by deep space tracking networks (\emph{e.g.,} NASA's DSN and ESA's Estrack) are routinely used for navigation and science applications. By ``shadow tracking" the spacecraft signal, Earth-based radio telescopes involved in Planetary Radio Interferometry and Doppler Experiment (PRIDE) can provide open-loop Doppler tracking data when the dedicated deep space tracking facilities are operating in closed-loop mode only.}
  {We explain in detail the data processing pipeline, discuss the capabilities of the technique and its potential applications in planetary science.}
   {We provide the formulation of the observed and computed values of the Doppler data in PRIDE tracking of spacecraft, and demonstrate the quality of the results using as a test case an experiment with ESA's Mars Express spacecraft.} 
   {We find that the Doppler residuals and the corresponding noise budget of the open-loop Doppler detections obtained with the PRIDE stations are comparable to the closed-loop Doppler detections obtained with the dedicated deep space tracking facilities.}
{}
   
   \titlerunning{Planetary Radio Interferometry and Doppler Experiment (PRIDE) Technique 2}
   \maketitle

%

\section{Introduction}

The Planetary Radio Interferometry and Doppler Experiment (PRIDE) technique exploits the radio (re-)transmitting capabilities of spacecraft from the most modern space science missions \citep{Duev2012}. A very high sensitivity of Earth-based radio telescopes involved in astronomical and geodetic Very Long Baseline Interferometry (VLBI) observations, as well as an outstanding signal stability of the radio systems allow PRIDE to conduct precise tracking of planetary spacecraft. The data from individual telescopes are processed both separately and jointly, to provide Doppler and VLBI observables, respectively. 
Although the main product of the PRIDE technique is the VLBI observables (\citet{Duev2016}, paper 1 of this series), the accurate examination of the changes in phase of the radio signal propagating from the spacecraft to each of the ground radio telescopes on Earth, make the open-loop Doppler observables derived from each telescope very useful for different fields of planetary research. 

Dedicated deep space tracking systems (\emph{e.g.,} NASA's Deep Space Network (DSN) and ESA's tracking station network (Estrack)) provide data to determine the precise state-vector of a spacecraft, based on the spacecraft signal detected at the ground-based receivers. For this end, the tracking systems can provide a variety of radiometric data (\emph{e.g.,} Doppler, range and interferometry data) under different operational schemes \citep{Thornton2003}. The type of tracking data needed for a particular spacecraft depends on which stage of its mission the spacecraft is at, and for which means these data will be used. However, this does not imply that several tracking data types cannot be used for the same purpose. In fact, the use of different precise and reliable tracking techniques not only enables a more challenging navigation performance, but could enhance various scientific experiment carried out during the mission \citep{MUR06,Mazarico2012,Iess2014}.

The Doppler effect due to the relative motion of the radio elements can be retrieved from the signal received at the ground station in different ways. One way is to measure the changes in light travel time of the received spacecraft signal with a closed-loop mechanism \citep{DSMS207}. In this tracking scheme, the received signal at the station is mixed with a local oscillator signal. Once the carrier frequency is found at the receiving station, a numerically controlled oscillator is set at the same value of the detected frequency and the carrier loop is closed \citep{Tausworthe1966,Gupta1975}. The bandwidth of the loop is gradually reduced to a pre-set operational value using its feedback mechanism. Once the `phase-lock' is acquired, the resulting Doppler shifted beat frequency is input into a Doppler cycle counter. The cycle counter measures the total phase change of the Doppler beat over a count interval, thus yielding the change in range over the count interval. The output - the Doppler cycle count-, consisting of an integer number from the Doppler counter itself and a fractional term from a Doppler resolver, is used to reconstruct the received spacecraft frequencies, also known as \textsl{sky frequencies} \citep{Morabito1995,Moyer2005}. The precision at which these measurements can be obtained, is limited by the way the time is tagged (\emph{i.e.,} the quality of the timing standards) and the Signal-to-Noise Ratio (SNR) of the measurements. Due to the mechanism used, the data derived is commonly known as \textsl{Doppler closed-loop} data.

The straightforwardness of this technique and the real-time availability of the data make closed-loop Doppler tracking the preferred tracking scheme when performing navigation and telemetry measurements with the DSN and Estrack networks. However, for radio science applications this is not necessarily the case. The term `radio science' includes all the scientific information that can be derived from the interaction of the spacecraft signal with planetary bodies and interplanetary media as it propagates from the spacecraft to Earth \citep{Tyler1989, Howard1992, Kliore2004, Patzold2004, Hausler2006, Iess2009}. In some scenarios, for instance planetary atmospheric occultation \citep{Jenkins1994,Tellmann2009,Tellmann2013} and ring occultation \citep{Marouf1986}, the received signal can present abrupt changes in frequency and amplitude, yielding a loss-of-lock in a closed-loop tracking scheme. For such cases, an open-loop receiver is preferable. In this case, no real-time signal detection mechanism is present, but instead the frequency spectrum of the detected signal is downconverted, digitized and recorded, with a sufficiently wide bandwidth to be able to capture the high-dynamics of the signal \citep{DSMS209}. The processing of the data is performed at a later stage, using digital phase-lock loop (PLL), which simulates the real-time PLL-controlled system used in the closed-loop receivers, and a fast Fourier transform (FFT) that estimates the frequency and amplitude of the received signal. The difference resides in the ability of the digital PLL of starting new locking processes once the system is considered out of lock, and the direct estimation of the frequency of the carrier tone at each sampling time. This mechanism allows an observer to directly reconstruct the sky frequency of parts of the detected signal that would be otherwise considered lost. For the post-processing of the open-loop Doppler, although it relies on the same main detection methods (PLL and FFT), there are different spectral analysis approaches that can be used \citep{Lipa1979, Tortora2002,Paik2011,Jian2009}.

We present our approach for deriving Doppler open-loop data with the PRIDE technique, using a set of radio telescopes from the European VLBI Network (EVN) and the Very Large Baseline Array (VLBA). Although these telescopes are typically used for observations of natural cosmic radio sources, ranging from nearby stars to distant quasars, we have demonstrated in the past that our approach, based on precision wideband spectral analysis, is capable of tracking planetary spacecraft signals \citep{OVERHUYGENS06,Duev2012,Molera2014,Duev2016}. Since their conception, the equipment and the data acquisition software of the DSN and VLBI networks have been developed in close collaboration between the two scientific communities. For this reason, the characteristics and capabilities of the VLBI network receivers and the DSN/Estrack open-loop receivers (also known as radio science receivers) are very similar. The post-processing techniques, however, may differ even between radio science teams using the same network, because once the data is recorded the tracking center delivers it to the science teams, who use their own software for data processing and analysis.

For these reasons, this paper has two goals: 1) To present our processing technique to derive the open-loop Doppler data, provide a clear formulation of the observed and computed Doppler observables, and a noise budget of the derived tracking data. In this way we will analyze the quality of open-loop Doppler data derived with VLBI telescopes through the PRIDE technique and compare it to the standards of the closed-loop Doppler data provided by the deep space networks.  2) That since PRIDE uses another network of ground stations, it allows for the possibility of acquiring precise Doppler open-loop data independently from the space agencies' tracking networks. For instance, one potential applicability is to use PRIDE, with the EVN and VLBA networks, to track spacecraft when there are only closed-loop tracking passes scheduled by the corresponding agency's tracking facilities, for navigation and telemetry purposes. In this way, by performing shadow tracking on the spacecraft signal PRIDE could allow radio science activities to be conducted in parallel. These goals are addressed in this paper in the framework of the PRIDE tracking of the ESA Mars Express (MEX) spacecraft during its flyby of Phobos in December 2013.

The MEX orbiter was launched in June 2 2003 and has been orbiting the red planet since December 2003 in a highly elliptical polar orbit, with $86^{\circ}$ inclination, periaerion of $\sim 300$\,km, apoaerion of $\sim 10100$\,km and an orbital period of 6.7\,h. Due to its highly valuable science return the mission has been extended six times beyond its nominal mission duration \citep{Chicarro2004}. MEX's Telemetry, Tracking and Command (TT\&C) subsystem operates in a two-way mode, receiving the transmitted uplink signal in X-band (7.1\,GHz) and providing coherent dual-frequency carrier downlinks, at X-band (8.4\,GHz) and S-band (2.3\,GHz), via the spacecraft's 1.8\,m High Gain Antenna (HGA) for all radio science operations of the Mars Express Radio Science Experiment (MaRS) team \citep{Patzold2004}. On the ground MaRS activities are supported by the 35-m ESA Estrack New Norcia (NNO) station and the 70-m NASA Deep Space Network (DSN) stations, all equipped with hydrogen masers as part of the frequency and timing systems. On December 29 2013, MEX performed a Phobos fly-by at a distance of $\sim 45$\,km from its surface. Under the European Satellite Partnership for Computing Ephemerides (ESPaCE) consortium an opportunity was offered to track the spacecraft with the PRIDE technique using VLBI stations, along side the customary Estrack and DSN stations. The tracking session lasted for 25 hours around the flyby event, using 31 VLBI stations around the world, also equipped with hydrogen masers as frequency standards, observing at X-band (channel starting at 8412\,MHz, recording bandwidth of 16\,MHz) in a three-way mode. The PRIDE setup for this particular tracking experiment is described in detail in \citet{Duev2016} (paper 1).

The paper is organized as follows. In Section \ref{sec:formulation}, the processing pipeline to extract the Doppler detections from the raw open-loop data is explained, as well as the formulation of the observed and computed values of the instantaneous Doppler observables. In Section \ref{sec:GR035}, the observations and the open-loop Doppler detections obtained from ESA's MEX spacecraft in December 2013 are discussed. The quality of the PRIDE Doppler detections is assessed by comparing the Doppler noise obtained by the multiple VLBI stations involved in the experiment with the noise of the Estrack and DSN stations during the same tracking session. The main contributing noise sources are quantitatively discussed. Section \ref{sec:conclusions} summarizes the results and discusses how the findings can improve the planning and enhance the science return of future radio science experiments with PRIDE.

\section{PRIDE Doppler observables}
\label{sec:formulation}

In the nominal MEX gravimetry experiments, the orbit perturbations caused by the gravitational fields of Mars and - in this particular case - of Phobos are determined via precise two-way radio Doppler tracking of MEX with dual-frequency downlink during pericenter passes, with the Estrack and DSN stations \citep{Hagermann2009}. However, for the MEX Phobos fly-by on December 29, 2013, the PRIDE joined the tracking effort in a three-way mode (\emph{i.e.}, receiving the signal re-transmitted by the spacecraft with a network of radio telescopes of which none is the initial transmitting ground station, also known as ``shadow tracking"), in order to assess the performance of the technique.

\subsection{Observed values of the Doppler observables}
\label{sec:obsrevable}

The transmitting/receiving systems at DSN and Estrack stations used for spacecraft radiometric tracking can operate in a closed- or open-loop manner. In these networks, the primary receiver is the closed-loop receiver, which uses a mechanism to phase-lock onto the received carrier signal. In this setup, the receiver passband is continuously aligned to the peak of the carrier tone and its bandwidth is gradually narrowed, allowing the retrieval of real-time tracking data and telemetry \citep{DSMS207}. The open-loop receivers on the other hand, do not have such a feedback mechanism \citep{DSMS209}, hence the bandwidth of the receiver passband is predefined and remains fixed during each observation. For this reason, the carrier signal filtering and tracking is performed at a later stage using the accurately timed detection of the signal recorded at the ground station. The radio telescopes used in PRIDE only operate in the open-loop mode. At each station, the received signals are amplified, heterodyned to the baseband, digitized, time-tagged and recorded onto disks using the standard VLBI data acquisition systems, with Mark5 A/B or FlexBuff recording systems \citep{Lindqvist2014}. For the data processing, the disks can be shipped or the data are transferred directly via high-speed networks to the VLBI data processing center at the Joint Institute for VLBI ERIC (JIVE) in the Netherlands. 

\begin{figure}[!htp]
  \centering
  \subfloat[]
  {\includegraphics[width= 8 cm]{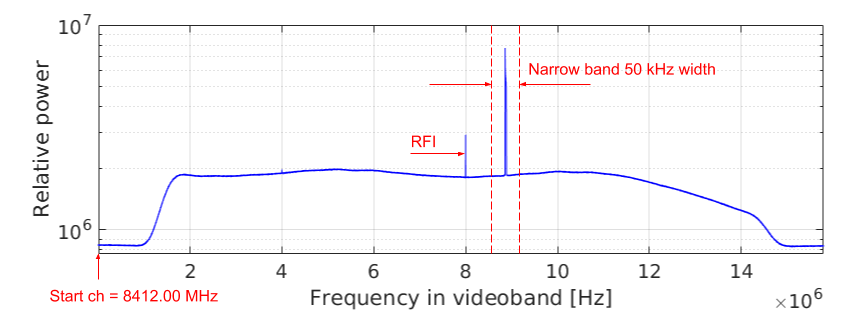}}\\     
  \subfloat[]
  {\includegraphics[width= 7.9 cm]{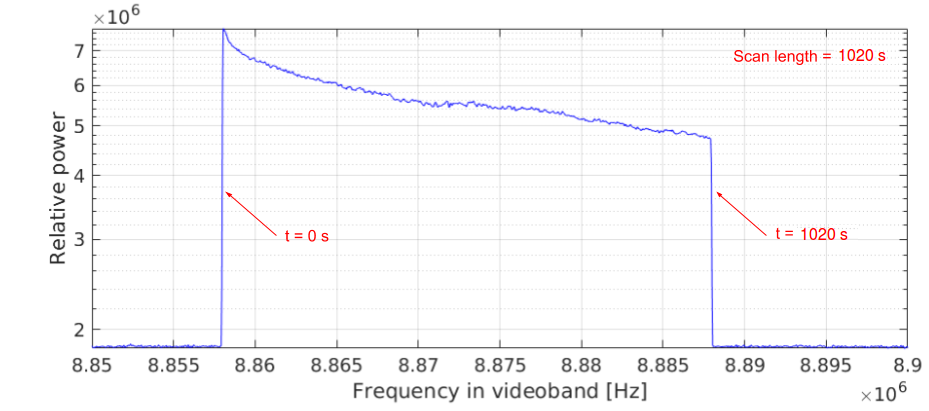}}\\
  \subfloat[]
  {\includegraphics[width= 8 cm]{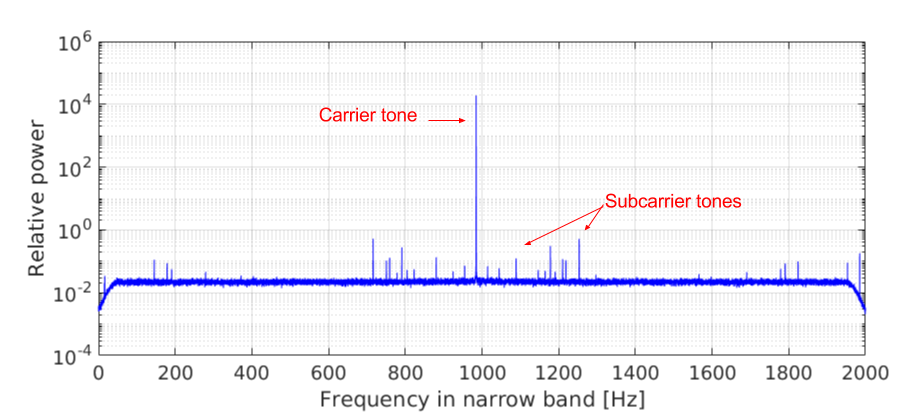}}\\ 
  \subfloat[]
  {\includegraphics[width= 8 cm]{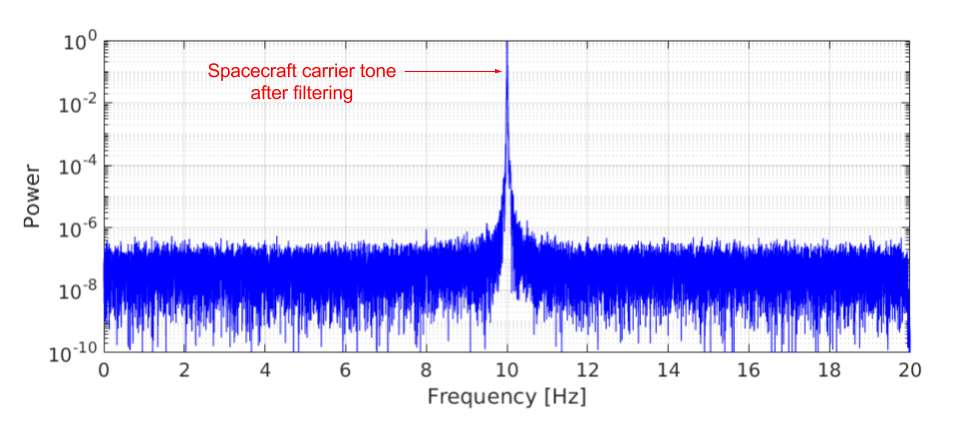}}\\

\caption[Doppler data processing pipeline.]{An example of Doppler data processing pipeline using observations of MEX during Phobos fly-by by Hartebeesthoek (see Section \ref{sec:GR035}). (a) is the typical resulting average power spectrum of a scan after running the \texttt{SWSpec} software. In the 16\,MHz pre-defined sub-band starting at 8412 MHz, the spacecraft signal is found in the spectrum. A narrow band containing the moving phase of the spacecraft carrier/tone is selected (in this case of 50\,kHz bandwidth) in order to model the Doppler shift using an $n$-order polynomial frequency fit. (b) shows a zoom of the spectrum inside the selected narrow band window to perform the fit. Here the moving phase of the carrier tone is visible along the duration of the scan. After the fit is performed, \texttt{SCtracker} applies the polynomial coefficients after converting the sample to baseband sample to stop the moving phase of the tone. In this way, \texttt{SCtracker} extracts an initial fit of the Doppler shift. (c) shows the output of the \texttt{SCtracker} a phase-stopped filtered out signal in a 2\,kHz narrow band in baseband. Finally, the \texttt{dPLL} performs high-precision iterations of the time-integration of the overlapped spectra, phase polynomial fitting, conversion to baseband and phase-stopping corrections, using narrowband windows around the carrier tone. The iterations stop when the window bandwidth reaches 20\,Hz, as shown in (d), allowing the extraction of the frequency and phase residuals of the spacecraft carrier tone with a 2\,mHz frequency spectral resolution.} 
\label{fig:scsoft}
\end{figure}

The Doppler detections are extracted from the raw open-loop data using the PRIDE spacecraft tracking software, consisting of three packages \texttt{SWSpec}, \texttt{SCtracker} and \texttt{dPLL}\footnote{https://bitbucket.org/spacevlbi/}\citep{Molera2014}. With the PRIDE setup we observe two sources, the spacecraft signal and natural radio sources that are used as calibrators. A large number of the natural sources observed with radio telescopes emit broadband electromagnetic radiation spanning many gigahertz in the frequency domain, however the signal is typically weak. It is therefore desirable to use as wide a frequency band as possible in order to detect the signal. The open-loop receiver systems of the VLBI stations are typically set up to record 4, 8, 16, or 32 frequency channels, with 4, 8, 16, or 32\,MHz bandwidth per sub-band. However, the spacecraft signal spectrum takes up only a fraction of the sub-band (see Figure \ref{fig:scsoft}a). 
For this reason, the first processing step is to extract the narrow band containing the spacecraft signal carrier and/or tones present in the spectrum. \texttt{SWSpec} extracts the data from the channel where the spacecraft signal is located, and subsequently performs a Window-OverLapped Add (WOLA) Direct Fourier Transform (DFT), followed by a time integration over the obtained spectra. The result is an initial estimate of the spacecraft carrier tone along the observation scan (Figure \ref{fig:scsoft}b). As shown in Figure \ref{fig:scsoft}b, the detected carrier tone has a moving phase throughout the scan, which is caused by the change in relative velocity between the spacecraft and the receiver. The goal is to extract the Doppler shift, first by fitting the changing frequency of the carrier tone by a $n$-order polynomial, and then using the fit to stop the moving phase of the tone. The latter step is performed with the \texttt{SCtracker} software, which subsequently allows the tracking, filtering and extraction of the tone in a narrow band. Figure \ref{fig:scsoft}c shows the narrowband output signal of the \texttt{SCtracker}. At this point, the spacecraft signal is in a band of a few kHz bandwidth, in contrast to the initial $4-32$ MHz bandwidth sub-band. The final step is conducted by the digital Phase-Locked-Loop (\texttt{dPLL}) which performs high precision reiterations of the previous steps -- time-integration of the overlapped spectra, phase polynomial fitting and phase-stopping correction -- on the narrow band signal. After the phase-stopping correction, the power spectrum is accumulated for a selectable averaging interval. Using a frequency window around the tone, the maximum value of the accumulated spectrum is determined. The corresponding frequency of the peak of the spectrum is stored, using as the time tag the middle of the averaging interval. This procedure is conducted throughout the whole range of spectra. The output of the \texttt{dPLL} is the filtered down-converted signal (Figure \ref{fig:scsoft}d) and the final residual phase in the stopped band with respect to the initial phase polynomial fit. The bandwidth of the output detections is typically about $20$ Hz with a frequency spectral resolution of $\sim2$\,mHz \citep{Molera2012}.

The PRIDE post-processing pipeline allows us to determine the instantaneous Doppler shift of the recorded tracking data, which is different from the integrated Doppler observables that are derived from closed-loop tracking data. For the purposes of orbit determination and the estimation of physical parameters of a celestial body using the Doppler data, it is important that this difference is taken into account when defining the observed and computed values of the Doppler observable. In the closed-loop case, the Doppler observables are derived by computing the change in the accumulated Doppler cycle counts from the spacecraft carrier phase measurements over a time interval at the receiver, as explained in detail in Section 13.3 of \citet{Moyer2005}. The corresponding modeled values are obtained by taking the difference in range at the beginning and end of the time interval. In the open-loop case, performed by PRIDE as explained in the previous paragraph, the observed values of the instantaneous Doppler observable (for one-way or three-way mode) are derived directly from an estimate of the carrier tone frequency of the spacecraft spectrum. Therefore, the observables are simply retrieved by adding the base frequency $f_{base}$ (which for the experiment analyzed in this paper was 8412\,MHz, as shown in Figure \ref{fig:scsoft}a) of the channel containing the spacecraft signal and the obtained time averaged tone frequencies $f_{tone}$ at each sampled time $t_i$,

\begin{equation}
f_R(t_i)= f_{base} + f_{tone}(t_i)
\end{equation}

where $f_R$ is the received frequency.

The integration time is defined by the number of FFT points used at the \texttt{dPLL} on the $\sim$2\,kHz bandwidth signal (Figure \ref{fig:scsoft}c). For the gravity field determination experiments, the desired integration time is $\sim$10\,s, hence 20,000 FFT points are used in \texttt{dPLL}. The uncertainty of each tone frequency estimate is derived from the final residual phase of the \texttt{dPLL} output.

\subsection{Computed values of the Doppler observables}
\label{sec:predictions}

To process the Doppler data obtained with PRIDE, a model is required that provides the instantaneous Doppler shift $f_{R}/f_{T}$, where $f_{R}$ and $f_{T}$ denote the observed frequency of the received and transmitted electromagnetic signal, respectively. Fundamentally, this frequency ratio is obtained from:
\begin{align}
\frac{f_{R}}{f_{T}}=\frac{d\tau_{T}}{d\tau_{R}}=\left(\frac{d\tau}{dt}\right)_{T}\frac{dt_{T}}{dt_{R}}\left(\frac{dt}{d\tau}\right)_{R}\label{eq:frequencyRatioFundamental}
\end{align}
where $\tau$ and $t$ denote proper time of the observer and coordinate time, respectively. The $R$ and $T$ subscripts denote properties of receiver and transmitter. The coordinate times of transmission and reception are related via the light-time equation:
\begin{align}
t_{R}-t_{T}=\frac{1}{c}\left|\mathbf{x}_{R}(t_{R})-\mathbf{x}_{t}(t_{T})\right|+\Delta(t_{R},t_{T})\label{eq:fundLightTimeEquation}
\end{align}
with $\mathbf{x}_{R}(t)$ and $\mathbf{x}_{T}(t)$ being the barycentric positions of the receiver and transmitter, and $\Delta(t_{R},t_{T})$ the relativistic correction to the light travel times.

The main complication in obtaining an explicit expression from Equation \ref{eq:frequencyRatioFundamental} is to compute the terms $d\Delta(t_{R},t_{T})/dt_{R}$ and $d\Delta(t_{R},t_{T})/dt_{T}$. To expand these equations, we use the formalism of \citet{Kopeikin1999}, where it is assumed that:
\begin{itemize}
\item The metric $g_{\alpha\beta}$ can be expanded to post-Minkowskian order, so that $g_{\alpha\beta}(\mathbf{x},t)=\eta_{\alpha\beta}$ + $h_{\alpha\beta}(\mathbf{x},t)$, with $\eta_{\alpha\beta}$ the Minkowski metric and the metric perturbation $h_{\alpha\beta}=O(G)$.
\item All bodies with gravity fields that perturb the null geodesic of the electromagnetic signal can be modeled as point masses.
\item All bodies with gravity fields that perturb the null geodesic of the electromagnetic signal have a constant barycentric velocity over the relevant time interval of a single measurement.
\end{itemize}
Under these assumptions, $\Delta(t_R, t_T)$ reduces to the following, neglecting second order terms in $v/c$:
\begin{align}
\Delta(t_R,t_T) & = -\frac{2G}{c^3} \sum_{a=1}^N m_a \left ( \ln{ \frac{r_a(t_R,s_R)-{\bf k}\cdot{\bf r}_a(t_R, s_R)}{ r_a(t_{T},s_{T})-{\bf k}\cdot{\bf r}_a(t_{T},s_{T}) } } \right. \nonumber \\
                  & - \left( {\bf k} \cdot \frac{{\bf v}_a(t_R, s_R)}{c} \right) \ln{( r_a(t_R, s_R) - {\bf k}\cdot{\bf r}_a(t_R, s_R) )} \nonumber \\
                  & \left. + \left( {\bf k} \cdot \frac{{\bf v}_a(t_{T}, s_{T})}{c} \right) \ln{( r_a(t_{T}, s_{T}) - {\bf k}\cdot{\bf r}_a(t_{T}, s_{T}) )} \right )
\end{align}
the static case of which is also known as the Shapiro effect \citep{Shapiro1971}. In the above, the parameter $s$ denotes the retarded time of body $a$ w.r.t. either the signal transmission or signal reception (for $T$ and $R$ subscripts, respectively). This time parameter is obtained from the light-time equation of Equation \ref{eq:fundLightTimeEquation}, only now by considering the perturbing body $a$ as the transmitting body, so that:
\begin{align}
t_{R}-s_{R}=\frac{1}{c}\left|\mathbf{x_{R}}(t_{R})-\mathbf{x_{a}}(s_{R})\right|\label{eq:lightTimeRetardedR}\\
t_{T}-s_{T}=\frac{1}{c}\left|\mathbf{x_{T}}(t_{T})-\mathbf{x_{a}}(s_{T})\right|\label{eq:lightTimeRetardedT}
\end{align}
Physically, these times represent the time at which the gravitational signal from body $a$ must be evaluated for its effect on the transmitter at $t_{T}$ and the receiver at $t_{R}$ to be modeled, implicitly assuming $c$ to be the speed of gravity. Note that although we omit any $a$ sub/superscript of the times $s$, we stress that these times are \emph{different} for each perturbing body $a$.

Under these above assumptions, as shown by \citet{Kopeikin1999}, Equation \ref{eq:frequencyRatioFundamental} can be written as:
\begin{equation}
f_R = f_T \frac{1 - {\bf k} \cdot {\bf v}_R/c }{1 - {\bf k} \cdot {\bf v}_T/c} R(\mathbf{\mathbf{v}_{R},\mathbf{v_{T},t_{R},t_{T}}})\label{eq:fRfT}
\end{equation}
where ${\bf v}_R$  and ${\bf v}_T$ are the barycentric velocity vector of the receiving station at reception time $t_R$ and at transmission time $t_T$, respectively. The term $R$ denotes a set of (special and general) relativistic corrections. The unit vector ${\bf k}$ is the direction along which the radio wave propagates at past null infinity (\emph{i.e.} when following the signal back along the null geodesic to $t\rightarrow -\infty$), which can be expressed as:

\begin{equation}
{\bf k}  = - {\bf K} -  \boldsymbol{\beta}(t_R, s_R) + \boldsymbol{\beta}(t_T, s_T) 
\label{eq:eqk}
\end{equation}  
where ${\bf K}$ is the geometric direction of the propagation of the electromagnetic wave in a flat space-time and $\boldsymbol{\beta}(t_R, s_R)$ and $\boldsymbol{\beta}(t_T, s_T)$ are the relativistic corrections as a function of the states of body $a$ at the retarded times of reception and transmission of the electromagnetic signal, respectively. These vectors are defined as follows,

\begin{strip}
\begin{align}
{\bf K}  &= - \frac{{\bf x}_R - {\bf x}_T}{|{\bf x}_R - {\bf x}_T|}  \label{eq:eqK}\\
\beta^i(t,s) & = - \frac{2G}{|{\bf x}_T - {\bf x}_R| c^2} \sum_{a=1}^{N} m_a \left[ \frac{ 1 - {\bf k} \cdot {\bf v}_a(s)/c }{ \sqrt{1 - v^2_a(s)/c^2}} \frac{r^i_a(t,s) - k^i ( {\bf k} \cdot {\bf r}_a(t,s) ) }{ r_a(t,s) - {\bf k}\cdot{\bf r}_a(t,s) } \right] \nonumber \\
               & - \frac{4G}{|{\bf x}_T - {\bf x}_R| c^2} \sum_{a=1}^{N}  \left [ \frac{ m_a }{ \sqrt{1 - v^2_a(s)/c^2}} \left[ v^i_a(s)/c - k^i ( {\bf k} \cdot {\bf v}_a(s)/c ) \right ] \ln{ \left( r_a(t,s) - {\bf k} \cdot {\bf r}_a(t,s) \right)}  \right ]
\label{eq:betaK}
\end{align}
\end{strip}

The relativistic term $R$ in Equation \ref{eq:fRfT} can be decomposed as follows:
\begin{equation}
R ( v_R, v_T, t_R, t_T)= \left[ \frac{ 1 - (v_T/c)^2}{ 1 - (v_R/c)^2} \right]^{1/2} \left[ \frac{a(t_T)}{ a(t_R)} \right]^{1/2} \frac{b(t_R)}{b(t_T)}
\end{equation}
where the first term accounts for the special relativistic Doppler shift, the second term accounts for the general relativistic corrections due to the $d\tau/dt$ terms in Equation \ref{eq:frequencyRatioFundamental}, and the final term is (along with the terms $\boldsymbol{\beta}$ given above) a result of expanding $d\Delta/dt$ when inserting Equation \ref{eq:fundLightTimeEquation} into the middle term on the right-hand side of Equation \ref{eq:frequencyRatioFundamental}. The terms $a$ and $b$ are given by:
\begin{align}
a(t) & = 1 + \frac{2G}{c^2} \sum_{a=1}^N \frac{ m_a \sqrt{1 - v_a^2(s)/c^2}}{r_a(t,s) - {\bf v}_a(s) \cdot {\bf r}_a(t,s)/c } \nonumber \\
     & - \frac{4G}{c^2 - v^2} \sum_{a=1}^N \frac{m_a}{\sqrt{1 - v_a(s)^2/c^2}} \frac{(1 - {\bf v}(t) \cdot {\bf v}_a(s)/c^2 )^2}{ r_a(t,s) - {\bf v}_a(s)\cdot {\bf r}_a(t,s)/c}
\end{align}

\begin{align}
b(t) & = 1 + \frac{2G}{c^2} \sum_{a=1}^N \frac{ m_a }{ \sqrt{1 - v_a^2(s)/c^2} } \frac{ 1 - {\bf k }\cdot{\bf v}_a(s)/c}{ r_a(t,s) - {\bf v}_a(s)\cdot {\bf r}_a(t,s)/c} \cdot \nonumber \\
     & \left [ \frac{( 1 - {\bf k} \cdot {\bf v}_a(s)/c)({\bf k} \times {\bf v}(t)/c)\cdot({\bf k} \times {\bf r}_a(t,s))}{ r_a(t,s) - {\bf k}\cdot{\bf r}_a(t,s) } \right. \nonumber \\
     & \left. - \frac{({\bf k} \times {\bf v}_a(s)/c)\cdot({\bf k} \times {\bf r}_a(t,s))}{ r_a(t,s) - {\bf k}\cdot{\bf r}_a(t,s) } + {\bf k}\cdot {\bf v}_a(s)/c \right ]
\end{align}

Evaluating this algorithm for the one-way Doppler case requires $2N+1$ solutions of light-time equations, once for Equation \ref{eq:fundLightTimeEquation} and $N$ for both Equations \ref{eq:lightTimeRetardedR} and \ref{eq:lightTimeRetardedT}, with $N$ the number of bodies perturbing the path of the signal. These equations are implicit and must be solved iteratively. For Equation \ref{eq:fundLightTimeEquation}, we must compute $t_{T}$ from a given $t_{R}$. We initialize $t_{T(1)}=t_{R}$, and iterate to find $t_{T(n+1)}$ from $t_{T(n)}$ using the Newton-Raphson method:
\begin{equation}
t_{T(n+1)} = t_{T(n)} - \frac{ t_R - t_{T(n)} - \frac{|{\bf x}_R - {\bf x}_T|}{c} - \Delta_{(n)}( t_R, t_{T(n)} )}{ \frac{ {\bf x}_R - {\bf x}_T}{|{\bf x}_R - {\bf x}_T|} \frac{{\bf v}_{T}(t_{T(n)})}{c} - 1}
\end{equation}
The iterative procedure converges when $| t_{T(n+1)} - t_{T(n)}| \leq \epsilon$ for some predefined small $\epsilon$. For the evaluation of the term $\Delta_{n}$, the unit vector ${\bf k}$ is also updated iteratively using Equations \ref{eq:eqk}, \ref{eq:eqK} and \ref{eq:betaK}, initially assuming ${\bf k}_{(1)} = - {\bf K}$.

After the convergence of $t_T$, the values for ${\bf k}$, ${\bf r}_a(t_T, s_T)$, ${\bf v}_a(t_T)$, ${\bf x}(t_T)$ and ${\bf v}(t_T)$ are  computed, and the values for the general relativistic corrections $a(t_T)$, $a(t_R)$, $b(t_T)$ and $b(t_R)$ are determined. Finally, the instantaneous one-way Doppler frequency at reception time $t_R$ can be determined from Equation \ref{eq:fRfT}. 

The explicit formula for the two/three-way observables, for the propagation of the radio signal emitted from the ground station on Earth, with position ${\bf x_T}$ and velocity ${\bf v_T}$ at $t_T$, then received and transponded back to Earth by a spacecraft with position ${\bf x_S}$ and velocity ${\bf v_S}$ at $t_S$, where the superscripts `+' and `-' denote received on uplink and transponded on downlink, and finally received at a ground station on Earth, with position ${\bf x_R}$ and velocity ${\bf v_R}$ at $t_R$, is,

\begin{equation}
f_R = f_T M \left( \frac{1 - {\bf k}^+ \cdot {\bf v}_S^+ }{1 - {\bf k^+} \cdot {\bf v}_T}  R( v_T, v_S^+, t_T, t_S^+) \right) \frac{1 - {\bf k}^- \cdot {\bf v}_R }{1 - {\bf k^-} \cdot {\bf v}_S^-}  R( v_S^-, v_R, t_S^-, t_R)
\label{eq:fRfT2}
\end{equation}
where again all the positions and velocities are given with respect to the Solar System barycenter.

The solution to Equation \ref{eq:fRfT2} (the three-way Doppler predictions) is found in a similar manner as for Equation \ref{eq:fRfT} (the one-way predictions), by first solving the expression for the uplink (in parenthesis in Equation \ref{eq:fRfT2}), this time estimating the signal reception time at the spacecraft $t_S^+$ and subsequently determining all the uplink parameters corresponding to $t_S^+$ with the same iteration procedure as explained above. Once the expression in the parenthesis is solved, the downlink part is found as for Equation \ref{eq:fRfT2}, estimating the signal transmission time at the spacecraft $t_S^-$ and determining the corresponding downlink parameters.
In Equation \ref{eq:fRfT2}, $f_T$ is the frequency transmitted by the ground station at time $t_T$ and $M$ is the corresponding spacecraft turnaround ratio. The terms $R^+$ and $R^-$ are the relativistic corrections on uplink and downlink, respectively.

We note that in a practical implementation of this algorithm, it is important to explicitly recompute all state vectors at each step, as any expansion quickly becomes inaccurate on typical deep-space craft light travel times.

\section{MEX Phobos Fly-by: \texttt{GR035} experiment}
\label{sec:GR035}

The global VLBI experiment was designed to track MEX 14 hours prior to and 11 hours after its closest-ever Phobos fly-by at approximately 7:21 (UTC) on December 29, 2013. At the time of the experiment, Mars was at a distance of $\sim1.4$\,AU from the Earth with a Solar elongation of $\sim\,87$\,degrees. During the 25 hours, MEX was tracked by the Estrack New Norcia (NNO, western Australia) station, DSN DSS-63 (Robledo, Spain) and DSS-14 (Goldstone, California, USA) stations, and 31 VLBI radio telescopes around the world. The latter have been organized through the global VLBI experiment \texttt{GR035}. The experimental setup of \texttt{GR035} has been presented in \citet{Duev2016}. During the first 9 hours, NNO was the transmission station, followed by 8 hours of tracking with DSS-63 and finally 8 hours with DSS-14. The distribution of the telescopes over the duration of the experiment is presented in Figure 2, 3 and 7 in \citet{Duev2016}. The spacecraft operated in the two-way mode with an X-band uplink (7.1 GHz) and dual simultaneous S/X-band downlink (2.3/8.4 GHz) transponded by the High Gain Antenna (HGA) pointing towards the Earth. The Estrack and DSN stations produced two-way S- and X-band Doppler closed-loop data products. From the 31 VLBI stations involved in the experiment, only the detections of 25 stations are used for the analysis presented in this paper, as listed in Table \ref{tab:GR035stationsinfo}, because of different technical problems with the remaining 6 stations during the observation. 

\begin{figure}[!htp]
\centering
\includegraphics[width=9.7cm]{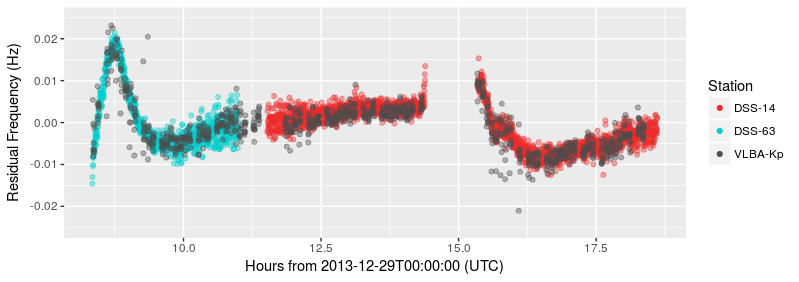}
\caption{Comparison of the Doppler residuals obtained with VLBA-Kp (in black), DSS-63 (in blue) and DSS-14 (in red). The median value of the difference between the fit of VLBA-Kp and the fit of DSS-63 and DSS-14, respectively, remains below 1\,mHz, for an integration time of 10\,s.}
\label{fig:residuals}
\end{figure}

We formed the Doppler residuals by differencing the Doppler detections, obtained as explained in Section \ref{sec:obsrevable}, with the predicted Doppler, derived as explained in Section \ref{sec:predictions} using the latest MEX navigation post-fit orbit of December 2013 provided by the European Space Operations Centre (ESOC)\footnote{ftp://ssols01.esac.esa.int/pub/data/ESOC/MEX/ORMM\_FDLMMA\_ DA\_131201000000\_01033.MEX}. In order to correctly determine the Doppler noise of the residuals, we flagged some data out. In particular, the data obtained during occultations and the fly-by event were discarded. Additionally, we discarded a number of scans that presented systematic outliers, for instance, when the transmission mode at the ground station changed. 
We found that the Doppler residuals obtained with the VLBI stations are in agreement with the DSN and Estrack residuals. Figure \ref{fig:residuals} shows as an example the frequency residuals found with the 25-m antenna of the Very Long Baseline Array (VLBA) at Kitt Peak (designation Kp) and the residuals of the 70-m DSS-63 and DSS-14 antennas. In this case, the median value of the difference between the fit of Kp and the fit of DSS-63 and DSS-14, respectively, remains below 1\,mHz for an integration time of 10\,s. For other VLBI stations, the median (after flagging) of the Doppler residuals found to be approximately 2\,mHz. 

In order to determine the quality of the PRIDE Doppler detections of the different VLBI stations involved in this experiment, first we have to identify the different sources that contribute to the overall noise of the Doppler residuals. The signal received at the ground stations will have random errors introduced by the instrumentation on board the spacecraft and at the receiving system, and the random errors introduced by the propagation of the signal through the different media along the line of sight of the ground station. Additionally, systematic errors can also be introduced, for instance, when calibrating the signal or in the models used to generate the predicted Doppler signature. The calibration of the Doppler observables in relation to the signal delays induced by the Earth's ionosphere are performed using the total vertical electron content (vTEC) maps available from the International GNSS Service (IGS) on a daily basis with a 2-hour temporal resolution on a global grid \citep{FELTENS09}. The calibration of the tropospheric signal delays is applied by using the Vienna Mapping Functions VMF1 \citep{BOEHM06} or by ray-tracing through the Numerical Weather Models (NWM) \citep{DUEV11}, depending on the antenna elevation. The systematic errors due to the model and orbit used to derive the Doppler residuals, as well as the residual systematic noise resulting from the ionospheric and tropospheric delay calibration, will not be characterized in this paper.

The random errors introduced by the instrumentation will be analyzed in Section \ref{sec:IntrumentalNoise}, and the random errors introduced by the propagation of the signal through the interplanetary media will be treated in Section \ref{sec:PropagationNoise}. Finally, in Section \ref{sec:NoiseBudget} the summary of the noise budget is given.

\subsection{Instrumental noise}
\label{sec:IntrumentalNoise}

The noise budget of the two-way Estrack/DSN Doppler detections and the three-way PRIDE Doppler detections will include instrumental noise introduced at the transmitting ground station (electronics, frequency standard, antenna mechanical noise) and at the spacecraft (electronics) that are common to both observables \citep{Asmar2005,Iess2014}. Hence, regarding instrumental noises, the difference between these noise budgets resides at the receiving stations: the thermal noise, induced by the ground station receiver system and the limited received downlink power, the frequency and timing systems' noise, and the antenna mechanical noise. In this paper, only the first two sources of noise will be treated since the antenna mechanical noise of the VLBI stations has not yet been characterized for the time intervals relevant to this study.

The thermal noise of the ground station is characterized by the RMS of the random fluctuations of the total system power at the ground station.
The one-sided phase noise spectral density $S_{\phi}$ of the received signal gives the relative noise power to the carrier tone, contained in a 1\,Hz bandwidth chosen to be centered at a frequency with a large offset $\Delta f$ from the carrier frequency $f_{carrier}$ \citep{Vig1999},

\begin{equation}
S_{\phi} = \frac{P_{sideband}(f_{carrier} + \Delta f)}{P_{carrier}}
\label{eq:sphi}
\end{equation}

where $P_{carrier}$ is the power of the carrier tone and $P_{sideband}$ is the power of the 1\,Hz bandwidth band.

The SNR is then approximated by $1/S_{\phi}$. As explained in \citep{Barnes1971,Rutman1991}, the Allan deviation of white phase noise can be estimated by,

\begin{equation}
\sigma_y(\tau) \approx \frac{\sqrt{ 3 \mbox{B} S_{\phi} } } { 2 \pi f_0 \tau}
\label{eq:sigmaySNR}
\end{equation}

Using Equations \ref{eq:sigmaySNR} and \ref{eq:sphi}, and since $\mbox{SNR}(\tau, \mbox{B}) = \mbox{SNR}( 1\,\mbox{s}, 1\,\mbox{Hz})\sqrt{\frac{\tau}{\mbox{B}}}$, the Allan deviation for the SNR detections of the different telescopes were determined, as shown in Table \ref{tab:GR035stationsinfo}. During the Phobos fly-by science operations conducted with MEX, the two-way closed-loop Doppler data is obtained using a carrier loop bandwidth of 30\,Hz at the ground stations, with 10\,s integration time \citep{Hagermann2009}. For the VLBI telescopes, the three-way open-loop Doppler data was initially recorded in a 16\,MHz wide band and then processed with the software described in Section \ref{sec:obsrevable}, for a final phase detection of 20\,Hz bandwidth, with 10\,s integration time. During the MEX orbits where NNO and DSS-14 were the transmitting stations, the telemetry was being transmitted except during MEX's nominal observation phase around the pericenter passage. However, during the orbit where DSS-63 was the transmitting station, in which the Phobos fly-by occurred, the telemetry was turned off throughout the whole orbit. For the Doppler error budget, only the time slots planned as radio science passes are taken into account, since the average SNR calculated with Equation \ref{eq:sphi}  drops when the telemetry is on. In Table \ref{tab:GR035stationsinfo} we also give the values of each telescope's sensitivity, which for single-dish radio telescopes is defined as the System Equivalent Flux Density (SEFD) \footnote{SEFD is equal to $2kT_{sys}/A_e$, where $A_e$ is the antenna's effective collecting area, $T_{sys}$ is the total system noise temperature and $k$ is the Boltzmann constant. The SEFD is a measurement of the performance of the antenna and receiving equipment since it gives the flux density (in Jy) produced by an amount of power equal to the off-source noise in an observation.}. This value is useful when comparing the performance between ground stations, since it comprises information about the total noise of the system and the collecting area of the antenna. Also when planning an experiment, it is important to know the nominal SEFD of a station at given frequency, since it can used to compute the expected SNR of a detection. 

\begin{figure}[!htp]
\centering
\includegraphics[width=9.7cm]{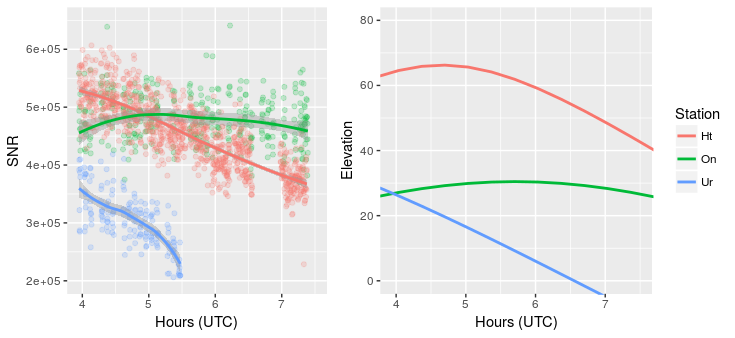}
\caption[SNR comparison between Ur, On, and Ht radio telescopes.]{SNR and elevation angle comparison between 25-m Ur, 25-m On and 15-m Ht radio telescopes from 4:00 - 7:30 UTC on December 29 2013 (TX: DSS-63). The left panel shows that although Ht (in pink) has a lower collecting area, it achieves higher SNR levels than Ur (in blue) and similar SNR levels than On (in green). The right shows the elevation angle of each antenna during the same time period, from which the correlation with the SNR levels of each station is evident.}
\label{fig:gr035DSS63OnHtUr}
\end{figure}

The 65-m dish Tianma \footnote{With exception of Tianma, the VLBI stations involved in the experiment have much smaller collecting area than the 70-m DSS-14 and DSS-63. However, there are several of the VLBI stations whose diameters are close to that of NNO (\emph{e.g.,} Ys, Sv, Zc, Bd and Km).} has the highest SNR detections, with a downlink power at reception 15\,dB higher than for the smallest stations, the 12-m Yarragadee, Katherine, Hobart and Warkworth - as it was expected because of their smaller collecting area - at elevations higher than 30 degrees. At $\tau = 10\,s$ the corresponding Allan deviation is of $4.6 \times 10^{-15}$. Table \ref{tab:GR035stationsinfo} uses the average SNR values over the whole coverage of each telescope. However, due to the large variation of the antennas' elevation during the several hours-long tracking sessions, there are periods of time for which smaller antennas achieve similar SNR levels as larger antennas. Figure \ref{fig:gr035DSS63OnHtUr} shows such an example in which for 3 hours the 15-m Hartbeesthoek achieves better SNR levels than the 25-m Urumuqi and similar SNR levels than 25-m Onsala, because of a more favorable antenna elevation (at elevations $<20$ degrees, several noise contributions at the antenna rapidly increase, such as the atmospheric and spillover noise). 

Regarding the noise induced by the frequency and timing systems, the Estrack, DSN, VLBA and EVN stations are all equipped with hydrogen masers frequency standards, which provide a stability better than $4<10^{-14}$ at $\tau=10$\,s ($<10^{-15}$ at $\tau=1000$\,s). Hence the noise contributions related to the frequency standard, should be on the same order of magnitude for the different networks.

\begin{table*}[!htbp]
\centering
\caption{Thermal noise of X-band Doppler detections of the VLBI stations during the \texttt{GR035} experiment.}
\footnotesize
\begin{tabular}{@{}l c c c c c c @{} } 
\toprule
\multicolumn{ 1}{c}{Observatories} & \multicolumn{ 1}{c}{Location} & \multicolumn{ 2}{c}{Telescope} & \multicolumn{ 1}{c}{Average $T_{sys}$} & \multicolumn{ 1}{c} {SEFD} & \multicolumn{ 1}{c} {Allan Deviation}  \\ 
\multicolumn{ 1}{l}{} & \multicolumn{ 1}{l}{} & Code & Diameter (m) & (K) & Jy & at 10 s \\ 
\midrule
DSN Goldstone 		& USA 		& DSS-14 & 70 	 & 20.6**     & 20**              &  $6.5 \times 10^{-15}$***\\ \midrule
DSN Robledo 		& Spain 	& DSS-63 & 70 	 & 20.6**  	& 20**		& $6.5 \times 10^{-15}$*** \\ \midrule
Estrack New Norcia 	& Australia 	& NNO 	&  35 	 & 60.8** 	& 40**		& $6.5 \times 10^{-15}$***\\ \midrule
Yebes			& Spain 	& Ys 	& 40 	 & 41 	&  200    	& $5.6 \times 10^{-15}$  \\ \midrule
Onsala 			& Sweden  	& On-60 & 20 	 & 62 	& 1240 	& $9.5 \times 10^{-15}$  \\ \midrule
Svetloe 		& Russia 	& Sv 	& 32 	 & 58*	& 200*     	& $1.1 \times 10^{-14}$ \\ \midrule
Zelenchukskaya 		& Russia 	& Zc 	& 32 	 & 30*	& 200*     	& $5.8 \times 10^{-15}$  \\ \midrule
Badary 			& Russia 	& Bd 	& 32 	 & 27*	& 200*     	& $8.2 \times 10^{-15}$  \\  \midrule
Hartebeesthoek 		& South Africa 	& Hh 	& 26 	 & 70 	& 875  	& $8.3 \times 10^{-15}$  \\ 
               		&   		& Ht 	& 15 	 & 44 	& 1260  & $1.0 \times 10^{-14}$  \\ \midrule
Tianma 			& China 	& Tm65 (T6) & 65 & 26 	& 48*  	& $4.6 \times 10^{-15}$  \\ \midrule
Urumuqi			& China 	& Ur 	& 25 	 & 86 	& 350* 	& $1.2 \times 10^{-14}$   \\ \midrule
Sheshan 		& China 	& Sh 	& 25 	 & 32 	& 800*     	& $8.6 \times 10^{-15}$\\ \midrule
Yamaguchi 		& Japan  	& Ym 	& 32 	 & 50*	& 106*  	& $8.9 \times 10^{-15}$   \\ \midrule 
Hobart 			& Australia 	& Ho 	& 26 	 & 68 	& 2500 	& $1.1 \times 10^{-14}$   \\ 
       			&           	& Hb 	& 12 	 & 87 	& 3500 	& $1.5 \times 10^{-14}$ \\ \midrule
Ceduna 			& Australia 	& Cd 	& 30 	 & 85*   &  600*  & $8.1 \times 10^{-15}$ \\ \midrule
Yarragadee 		& Australia 	& Yg 	& 12 	 & 96 	&  3500	& $1.6 \times 10^{-14}$  \\ \midrule
Katherine 		& Australia 	& Ke 	& 12 	 & 112 	&  3500	& $1.5 \times 10^{-14}$ \\ \midrule
Warkworth 		& New Zealand 	& Ww 	& 12 	 & 94	&  3500	& $1.6 \times 10^{-14}$ \\ \midrule
VLBA Owens Valley 	& USA  		& Ov 	& 25 	 & 35	& 300	& $8.5 \times 10^{-15}$   \\ \midrule
VLBA Kitt Peak 		& USA 		& Kp 	& 25 	 & 36	& 310	& $6.2 \times 10^{-15}$  \\ \midrule
VLBA Hancock 		& USA 		& Hn 	& 25 	 & 49	& 419	& $7.0 \times 10^{-15}$   \\ \midrule
VLBA Brewster 		& USA 		& Br 	& 25 	 & 41	& 352	& $2.5 \times 10^{-14}$ \\ \midrule
VLBA Mauna Kea 		& USA 		& Mk 	& 25 	 & 43	& 368	& $9.9 \times 10^{-15}$  \\ \midrule
VLBA St Croix 		& USA 		& Sc 	& 25 	 & 39	& 330	& $9.4 \times 10^{-15}$  \\ \midrule
VLBA Pie Town 		& USA 		& Pt 	& 25 	 & 27	& 313	& $7.1 \times 10^{-14}$ \\ \midrule
VLBA Fort Davis 	& USA 		& Fd 	& 25 	 & 36	& 309	& $6.2 \times 10^{-15}$   \\ \midrule
\bottomrule
\end{tabular}
\scriptsize \\
*Nominal values taken from the EVN status table II \citep{EVNSTATUS}. \\
** Nominal values taken from \citet{Stelzried2008} and \citet{Martin2004}.\\
*** Assuming a nominal value of the suppressed modulation Carrier-to-Noise Ratio (CNR) of 67 dBHz.
\label{tab:GR035stationsinfo}
\end{table*} 



\subsection{Medium propagation noise}
\label{sec:PropagationNoise}

The precision of the Doppler detections is also affected by the noise introduced by the propagation of the radio signal through the interplanetary medium, ionosphere and troposphere. The effects of ionospheric and interplanetary scintillation can be studied using the differenced phases of the signals received in S-band, $\phi_{s}$, and X-band, $\phi_{x}$, $\phi_{\Delta}(t) = \phi_s - \frac{3}{11}\phi_x$ \citep{Levy1977}. By subtracting the phases, the contribution of the dispersive plasma scintillations can be isolated. The Allan variance of the differenced phases is related to its two-sided phase power spectrum $S_{\phi}(f)$ \citep{Barnes1971,Armstrong1979} by

\begin{equation}
\sigma_y^2 (\tau) = \int^{\infty}_0 S_{\phi}(f) \frac{f^2}{\nu_0^2} \frac{sin^4(\pi \tau f)}{(\pi \tau f)^2} df
\label{eq:allanscint}
\end{equation} 

As explained in \citet{Armstrong1979}, when the phase spectrum can be approximated to $S_{\phi}(f) = A f^{-m}$, Equation \ref{eq:allanscint} can be rewritten as

\begin{equation}
\sigma_y^2(\tau) = \frac{A \tau^{m}}{\pi^2 \nu_0^2 \tau^3} \int^{\infty}_0 \frac{\sin{(\pi z)}^4}{z^m}dz
\label{eq:allanscinteasy}
\end{equation}

\begin{figure}[!htp]
\centering
\includegraphics[width=9 cm]{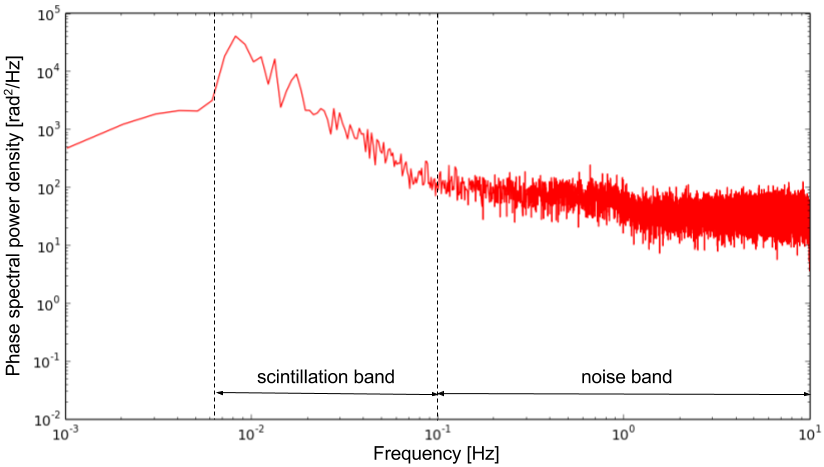}
\caption[Spectral power density of MEX signal in S-band.]{Spectral power density of MEX signal in S-band. The scintillation band extends from 8\,mHz to 0.1\,Hz, obtaining a value for the slope of -2.471, which is coherent with the spectral index values found by \citep{Woo1979}. The mean phase scintillation index, of the signal received from 3h56m to 09h12m (UTC) on 2013-12-29, is 0.070 rad, at an elongation of $\sim 87^{\circ}$ and distance of $\sim 1.4$\,AU.}
\label{fig:SpectralPowerDensitySband}
\end{figure}

\begin{figure}[!htp]
\centering
\includegraphics[width=9 cm]{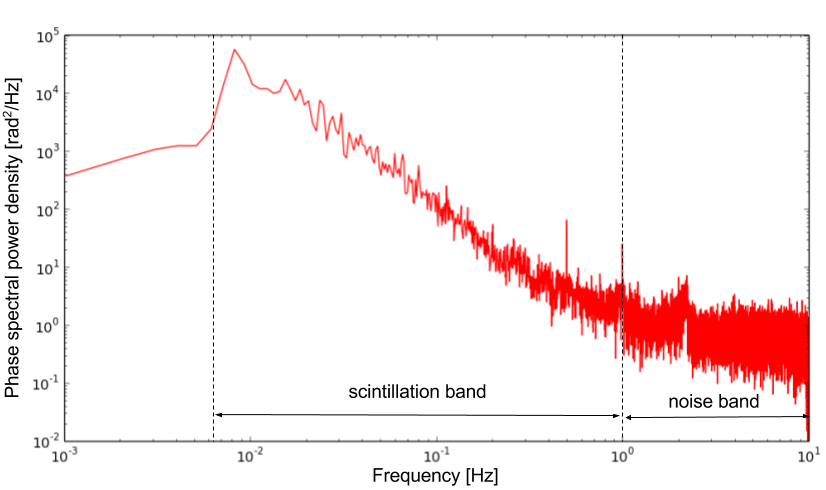}
\caption[Spectral power density of MEX signal in X-band.]{Spectral power density of MEX signal in X-band. The scintillation band extends from 8\,mHz to 0.45\,Hz, obtaining a value for the slope of -2.469. The mean phase scintillation index, of the signal received from 3h56m to 09h12m (UTC) on 2013-12-29, is 0.073 rad, at an elongation of $\sim 87^{\circ}$ and distance of $\sim 1.4$\,AU.}
\label{fig:SpectralPowerDensityXband}
\end{figure}

A first-order approximation of the phase spectrum on a logarithmic scale is performed as explained in \citep{Molera2014}, from which the slope $m$ and the constant $A$ are determined using only the Doppler-mode\footnote{In the Doppler-mode the telescopes observe in a dual S/X-band (2/8.4 GHz) frequency set-up, pointing iteratively at the spacecraft for 20 minutes and then 2 minutes at the calibrator, as explained in \citet{Duev2016}.} observations, where the length of the scan is typically $>10$ minutes. For instance, considering the Doppler-mode observations of Hartebeesthoek 15-m antenna, the plasma phase scintillation noise can be characterized. Figures \ref{fig:SpectralPowerDensitySband} and \ref{fig:SpectralPowerDensityXband} show the spectral power density of the phase fluctuations of MEX signal in S- and X-band, respectively. The lower and upper limits of the scintillation band are determined by visual inspection, taking into account the cut-off frequency defined to perform the polynomial fit and the amount of fluctuation due to the receiver system noise. The phase scintillation indices obtained with the S- and X-band signal, 0.070\,rad and 0.073\,rad, correspond to the results for Mars-to-Earth total electron content (TEC) along the line of sight found by \citet{Molera2017}, where the dependence of the interplanetary phase scintillation on elongation was studied using various MEX observations. When comparing Figures \ref{fig:SpectralPowerDensitySband} and \ref{fig:SpectralPowerDensityXband}, it is noticeable that in the spectral power, the noise band in S-band is much higher than in X-band. This is due to a higher thermal noise of the receiver and larger presence of RFI in this band. 

\begin{figure}[!htp]
\centering
\includegraphics[width=9 cm]{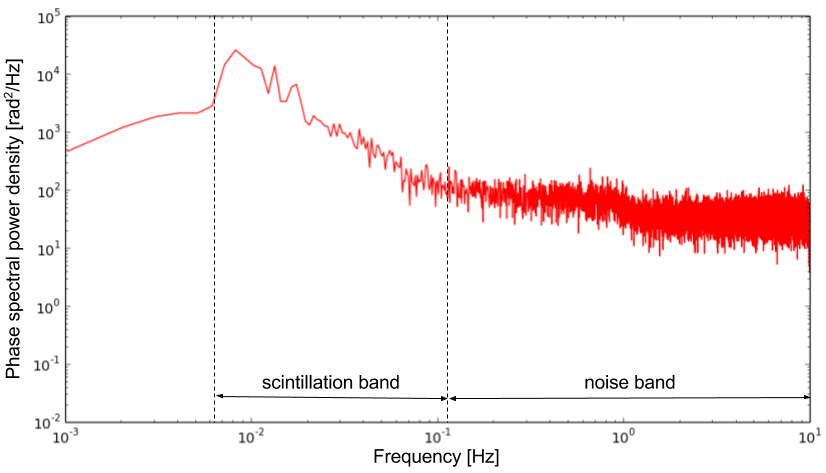}
\caption[Spectral power density of the differential phases.]{Spectral power density of the differential phases $\phi_{\Delta}$. The scintillation band extends from 8\,mHz to 0.1\,Hz, obtaining a value for the slope of -2.372. The mean phase scintillation index, of the signal received from 3h56m to 09h12m (UTC) on 2013-12-29, is 0.069 rad, at an elongation of $\sim 87^{\circ}$ and distance of $\sim 1.4$\,AU.}
\label{fig:SpectralPowerDensitySXband}
\end{figure}

Figure \ref{fig:SpectralPowerDensitySXband} shows the spectral power density of $\phi_{\Delta}$ for Ht. The slope found for the scintillation band that extends from 8\,mHz to 0.1\,Hz is -2.372 with a mean scintillation index of 0.069 rad. These results are in agreement with \citet{Molera2014}. As the phase power spectrum can be described in the form $S_{\phi}(f) = A f^{-m}$, following Equation \ref{eq:allanscinteasy} the Allan variance of the plasma phase scintillation is $2.46 \times 10^{-15}$ at $\tau=$1000\,s at $\sim 87^{\circ}$ elongation ( $4.44 \times 10^{-14}$ at $\tau=$10\,s).

More information regarding the origin of the phase fluctuations can be derived by analyzing the spatial statistics of the phase scintillation in multiple stations during the same tracking session. If the $\phi_{\Delta}$ phase data of a few pairs of widely spaced stations are cross-correlated, as suggested in \citet{Armstrong1998}, it could be determined whether the main contributor to the phase fluctuations is the interplanetary medium or the local impact of the ionosphere at each station. Unfortunately, in the \texttt{GR035} experiment this analysis could not be performed, since although the 3 stations operating in Doppler-mode are widely spaced (South Africa, Finland and China), the differential phase $\phi_{\Delta}$ could not be successfully retrieved due to high RFI on the S-band of Sh and technical problems with the X-band receiver of Mh (for this reason the values for Mh are not shown in Table \ref{tab:GR035stationsinfo}).

\subsection{Noise budget for the Doppler detections of \texttt{GR035}}
\label{sec:NoiseBudget}

Table \ref{tab:gr035NoiseBudget} summarizes the Allan deviations found for the different noise sources described in Sections \ref{sec:IntrumentalNoise} and \ref{sec:PropagationNoise}. The Allan deviations due to ground station's thermal noise for the VLBI stations vary from $\sigma_y(\tau) = 0.46 - 1.60 \times 10^{-14}$ at $\tau = 10$\,s. Despite the differences of the thermal noise $\sigma_y$ between the VLBI stations and the DSN and NNO stations \footnote{The Allan deviations for the DSN and NNO stations were calculated assuming an expected CNR suppressed modulation of 67 dB/Hz, as given in the level 1 data.}, the thermal noise of the stations do not dominate the error budget of the observations in this experiment, as shown in Table \ref{tab:gr035NoiseBudget}. It is worth mentioning, that due to the long tracking sessions of the antennas during this experiment, the antenna elevations have a higher impact in the detections' SNR than the collecting area of the antennas. This issue is usually ignored in shorter tracking sessions, since only stations with elevations $\sim >20$\,deg are selected to participate in an observation.

The plasma scintillation noise was estimated for Ht, which was one of the stations observing both in S- and X-band. The plasma scintillation noise is more dominant for Ht than its thermal noise ($\sigma_y = 4.44 \times 10^{-14}$ against $\sigma_y = 1.0 \times 10^{-14}$ for $\tau = 10$\,s). Due to problems with the receivers, this analysis could not be performed for the other two stations receiving the dual-band link. In future experiments, the contributions of the the ionosphere and interplanetary medium could be discerned from one another, by correlating the power spectra of the differential phases between every pair of stations.

\begin{table*}[!htbp]
\centering
\caption{Noise budget for PRIDE \texttt{GR035} experiment.}
\footnotesize
\begin{tabular}{@{}l c c @{} } 
\toprule
\multicolumn{ 1}{c}{Noise source} & \multicolumn{ 1}{c}{Allan deviation} & \multicolumn{1}{c} {Comments} \\
\multicolumn{ 1}{c}{ } & \multicolumn{ 1}{c}{at $\tau$=10\,s} & \multicolumn{ 1}{c} { } \\
\midrule

Ground station thermal noise & $0.5 - 1.5 \times 10^{-14}$ & For various sizes of antenna dishes (see Table \ref{tab:GR035stationsinfo}). \\ \midrule
Ground frequency reference source & $< 5.0 \times 10^{-14}$ & \citet{Tjoelker2010}  \\ \midrule
Plasma phase scintillation & $4.44 \times 10^{-14}$ & For Ht, at a solar elongation of $87^{\circ}$.  \\ \midrule
Antenna mechanical noise & -- & Has not been determined in this experiment. \\ \midrule

\bottomrule
\end{tabular}
\label{tab:gr035NoiseBudget}
\end{table*}

\citet{Armstrong2008} reported that for the DSN stations, when the propagation noises are properly calibrated, the antenna mechanical noise was the leading noise of their noise budget. Regarding the VLBI stations, \citet{Sarti2009,Sarti2011} have reported one-way path delay variations due to the antenna mechanical noise, however these were computed for VLBI geodetic and astrometric studies, for which the delay stability is evaluated in annual timescales, much larger than the integration times relevant for the study at hand. Nonetheless, due to their size \citep{Armstrong2016}, the expected mechanical noise of the VLBI antennas (except for Tm65) will be considerably less than the 70-m DSN antennas. In fact, simultaneous observations between PRIDE and DSN stations could help improve the sensitivity of the 70-m DSN antennas. Following the approach presented in \citet{Armstrong2008} in future experiments, stations of the global VLBI network close to the deep space tracking complexes could be used to remove the antenna mechanical noise of the larger antennas during simultaneous two-way/three-way Doppler passes, for instance, the 25-m VLBA-Ov close the DSS-14, the 14-m Ys close to the DSS-63, the 12-m Ye telescope close to NNO and the 12-m Atacama Pathfinder Experiment (APEX) telescope close to Estrack's Malarg\"{u}e station.

\section{Conclusions}
\label{sec:conclusions}

With the PRIDE setup, Doppler tracking of the spacecraft carrier signal with several Earth-based radio telescopes is performed, subsequently correlating the signals coming from the different telescopes, in a VLBI-style. Although the main output of this technique are VLBI observables, we have demonstrated that the residual frequencies obtained from the open-loop Doppler observables, which are inherently derived in the data processing pipeline to retrieve the VLBI observables, is comparable to that obtained with the closed-loop Doppler data from NNO, DSS-63 and DSS-14 stations (see Figure \ref{fig:residuals}). Figure \ref{fig:residuals} shows the best case found, where the median value of the residuals fit achieved with VLBI station Kp remains within 1\,mHz of the residuals fit obtained with DSS-63 and DSS-14. The median of the Doppler residuals for all the detections with the VLBI stations was found to be $\sim$ 2\,mHz.

The fact that this experiment involved long tracking sessions, makes the variability of the elevation angle of the antennas a factor in the characterization of the noise that cannot be ignored. At elevations <\,20\,degrees, noise contributions due to larger tropospheric path delays and larger spillover noise have a larger impact on the total system temperature compared to that of the receiver temperature. For this reason, there are cases for which antennas with smaller collecting areas reach similar SNR levels as larger antenna dishes, as shown in Figure \ref{fig:gr035DSS63OnHtUr}, due to a more favorable antenna elevation. The derived Allan deviations due to thermal noise at the VLBI stations vary between $\sigma_y(\tau) = 0.46 - 1.60 \times 10^{-14}$ at $\tau=10$\,s. For this particular experiment, at the DSN stations the expected $\sigma_y(\tau)$ due to thermal noise was $6.5 \times 10^{-15} $at $\tau$=10\,s. Although only 4 of the VLBI stations have comparable Allan deviations (Table \ref{tab:GR035stationsinfo}) to those of the DSN stations, the thermal noise is not the most dominant contribution to the overall noise budget of this experiment. 

It is important to mention that although they were not included in this particular experiment (only the 65-m Tianma station), PRIDE has through the EVN access to multiple radio telescopes similiar in size or larger than the DSN antennas, such as the 64-m Sardinia, 100-m Effelsberg and 305-m Arecibo, that can be scheduled for radio science experiments. The use of these large antennas can result in an advantage when conducting experiments with limited SNR, such as radio occultation experiments of planets/moons with thick atmospheres.   

Open-loop Doppler data, as those collected with PRIDE experiments, present advantages for certain radio science applications compared to closed-loop data. However, closed-loop Doppler tracking is routinely performed in the framework of navigation tracking and does not require post-processing in order to retrieve the Doppler observables.  Although the Estrack/DSN complexes have the capability of simultaneously gathering closed-loop and open-loop Doppler data, this is not an operational mode required for navigation nor telemetry passes, which generally operate in closed-loop mode only. In this sense, PRIDE Doppler data could complement the closed-loop tracking data and enhance the science return of tracking passes that are not initially designed for radio science experiments.

\begin{acknowledgements}
The EVN is a joint facility of European, Chinese, South African, and other radio astronomy institutes funded by their national research councils. The National Radio Astronomy Observatory is a facility of the National Science Foundation operated under cooperative agreement by Associated Universities, Inc. The Australia Telescope Compact Array is part of the Australia Telescope National Facility which is funded by the Commonwealth of Australia for operation as a National Facility managed by CSIRO.

T. Bocanegra Bahamon acknowledges the NWO–ShAO agreement on collaboration in VLBI.

G. Cim\'{o} acknowledges the EC FP7 project ESPaCE (grant agreement 263466).

P. Rosenblatt is financially supported by the Belgian PRODEX programme managed by the European Space Agency in collaboration with the Belgian Federal Science Policy Office.

We express gratitude to M. P\"{a}tzold (MEX MaRS PI) and B. H\"{a}usler for coordination
of MaRS and PRIDE tracking during the MEX/Phobos flyby and a number of valuable         comments on the manuscript of the current paper.

Mars Express is a mission of the European Space Agency.
The MEX a-priori orbit, Estrack and DSN tracking stations transmission frequencies and the cyclogram of events were supplied by the Mars Express project. The authors would like to thank the personnel
of the participating stations. R.M. Campbell, A. Keimpema, P. Boven (JIVE),
O. Witasse (ESA/ESTEC) and D. Titov (ESA/ESTEC) provided important support to various components
of the project.

The authors are grateful to the anonymous referee for the useful comments and suggestions.

\end{acknowledgements}

%
%

\bibliographystyle{bibtex/aa}
\bibliography{References/mybibliography_thesis}

\end{document}